\documentclass[aps,prb,twocolumn]{revtex4-1}
\usepackage{amsfonts}
\usepackage{amsmath}
\usepackage{amssymb}
\usepackage[pdftex]{graphicx}
\begin{document}
\title{Theory and simulations on strong pinning of vortex lines by nanoparticles}
\author{A. E. Koshelev}
\affiliation{Materials Science Division, Argonne National
Laboratory, Argonne, Illinois 60439, USA}
\author{A. B. Kolton}
\affiliation{CONICET, Centro At\'{o}mico Bariloche,
8400 S. C. de Bariloche, Argentina}

\date{\today }

\begin{abstract}
The pinning of vortex lines by an array of nanoparticles embedded inside
superconductors has become the most efficient practical way to achieve high
critical currents. In this situation pinning occurs via trapping of the
vortex-line segments and the critical current is determined by the typical
length of the trapped segments. To verify analytical estimates and develop a
quantitative description of strong pinning,  we numerically simulated isolated
vortex lines driven through an array of nanoparticles. We found that the
critical force grows roughly as the square root of the pin density and it is
strongly suppressed by thermal noise. The configurations of pinned lines are
strongly anisotropic, displacements in the drive directions are much larger
than in the transverse direction. Moreover, we found that the roughening index
for the longitudinal displacements exceeds one. This indicates that the local
stresses in the critical region increase with the total line length and the
elastic description breaks down in the thermodynamic limit. Thermal noise
reduces the anisotropy of displacements in the critical regions and straightens
the lines.
\end{abstract}
\pacs{74.25.Wx,74.25.Sv,74.20.De}
\maketitle

\section{Introduction}

The introduction of large-size nanoparticles of different shapes has
emerged as the best practical way to improve the current performance
of high-temperature superconductors. While in the first
superconducting cables the critical currents were limited by weak
links, in the second-generation superconducting wires based on
aligned YBa$_{2}$Cu$_{3}$O$_{7}$ (YBCO) films this problem has been
mostly resolved and critical currents are determined by vortex
pinning. Impressive progress has been made to enhance critical
currents in these films using both isotropic
\cite{Civale,HauganNat04,Song,Gutierrez,YamasakiSST08,MiuraPRB11,PolatPRB11}
and columnar \cite{MacManusNatMat04,Kang,VaranasiSUST07} inclusions.
In spite of this progress, our understanding of strong pinning
mechanisms is far from satisfactory. Theoretical estimates
describing the pinning of vortex lines by an array of strong pins at
low temperatures were elaborated by Ovchinnikov and
Ivlev.\cite{OvchinnIvlPRB91} This theory was applied to describe the
behavior of the critical currents in the real YBCO films in Refs.\
\onlinecite{BeekPRB02,IjaduolaPRB06}. In particular, frequently
observed power-law decay of the critical current as a function of
the magnetic field with a power slightly larger than $1/2$ is
naturally explained by this theory. More recently, it was also
argued that strong pins of unknown origin determine critical
currents at low magnetic fields in several iron pnictide
compounds.\cite{BeekPRB10} It is not clear, however, to what extent
available qualitative estimates describe the real situation. Due to
the obvious importance of strong pinning by large-size inclusions
for real superconducting materials, it is desirable to elaborate a
quantitative theory describing pinning in such situations. Moreover,
the very important issue of pinning suppression by thermal
fluctuations does not have any theoretical description in the
strong-pinning regime.

\begin{figure}[ht]
\begin{center}
\includegraphics[width=2.4in]{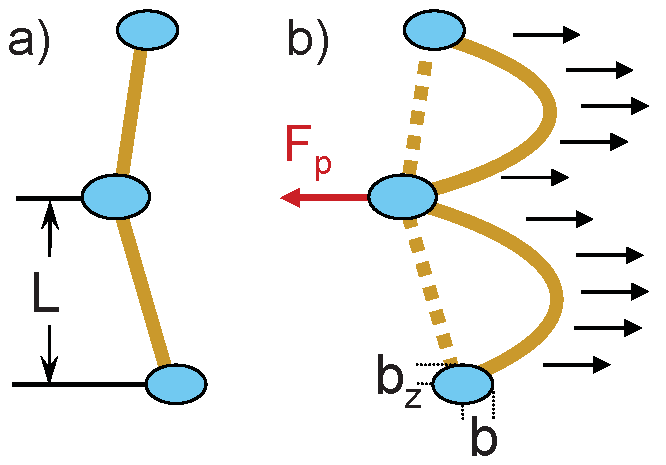}
\end{center}
\caption{Vortex line trapped by strong pinning centers at (a) zero
and (b) finite  current.} \label{Fig-TrappedLine}
\end{figure}
In this paper we consider the pinning of vortex lines in a
superconductor containing insulating inclusions with lateral sizes
larger than the coherence length. We focus on the pinning of
individual vortex lines corresponding to small magnetic fields.
Pinning occurs via the trapping of finite-size segments of a vortex
line\cite{OvchinnIvlPRB91} with a typical length $L$, as illustrated
in Fig. \ref{Fig-TrappedLine}. The critical current is determined by
the length of the trapped segment $L$ and the pin-breaking force
$F_{p}$,
\begin{equation}
\frac{\Phi_{0}}{c}j_{c}\approx\frac{F_{p}}{L}.
\label{CritCurrTrapSegm}
\end{equation}
Therefore, in the strong-pinning regime the critical-current problem
is mostly reduced to evaluation of the trapped-segment length $L$.
In general, trapping of the vortex lines is a complicated dynamic
process controlled by the competition between the pinning energy,
line tension, and intervortex interactions. Different approaches may
be used to evaluate the trapped-segment length. One can assume that
the pinning center always grabs a piece of the vortex line when it
is energetically favorable. This assumption implies that thermal
fluctuations facilitate local equilibration. In this case the
parameters of the trapped configurations can be obtained from the
energy-balance estimates. We will call this type of trapping the
equilibrium regime. This energy-balance consideration determines
trapped configurations of static vortex lines prepared by cooling in
finite magnetic field. However, such consideration is not applicable
to the more typical dynamic scenario when moving lines are trapped
after the driving force is slowly reduced down to the critical
value. In this case the line motion close to the critical force is a
continuous trapping-detrapping process limited by local
instabilities. The driven vortex can be captured when the line
either directly collides with a strong pin or passes sufficiently
close to it. In the second case trapping may occur due to the
long-range pin-vortex interaction as a result of local
instability.\cite{OvchinnIvlPRB91} The line remains trapped until
the force acting from the pin does not exceed the pin-breaking
force. When the line finally stops, the pinned configuration is
expected to be very anisotropic because the transverse pin-to-pin
displacements which are determined by trapping events are much
smaller than the longitudinal displacements limited by the
pin-breaking criteria.

To develop a quantitative picture and verify analytical estimates,
we explore in this paper the pinning of the vortex lines by
nanoparticles with extensive numerical simulations. We study the
dependence of the critical force on the density of pins, the
statistical properties of trapped lines including average values,
and the distributions of trapping length and pin-to-pin
displacements. We study long-range behavior of line displacements in
the direction of the driving force and in the transverse direction.
We also study in detail the suppression of the apparent critical
force by thermal fluctuations and the temperature dependence of the
trapping parameters.

The paper is organized as follows. In Section \ref{Sec-Inter} we describe
parameters characterizing the interaction between a vortex line and large-size
pinning centers. In Section \ref{Sec-AnalEst} we present analytical estimates.
This includes the formulation of general conditions for stable trapped
configurations in subsection \ref{SubSec-Cond}, making estimates for parameters
of trapped line in equilibrium, subsection \ref{SubSec-Equil}, and in the case
of dynamic trapping, subsection \ref{SubSec-DynTrap}. In the Section
\ref{Sec-NumModel} we describe the model used in our numerical simulations. In
the Section \ref{Sec-NumResults} we present our numerical results including the
zero-temperature case in subsection \ref{SubSec-zeroT} and the finite
temperature in subsection \ref{SubSec-finiteT} In Section \ref{Sec-Discussion}
we discuss our results and make preliminary comparisons with experiments.

\section{Interaction between vortex line and pinning center \label{Sec-Inter}}

We consider first the essential parameters describing the
interaction of vortex lines with large-size pinning centers. The
vortex pinning energy by an insulating spheroid inclusion with the
axes $b$ and $b_{z}$ is given by\cite{footnoteLp}
\begin{equation}
U_{p}\approx 2b_{z}\varepsilon_{0}\mathcal{L}_{p}, \ \
\mathcal{L}_{p} =\ln(b/\xi_{ab})\label{PinEnergy}
\end{equation}
with $\varepsilon_{0}\equiv \Phi_{0}^{2}/(4\pi\lambda_{ab})^{2}$.%
A very important parameter is the pin-breaking force, the maximum
force with which the pinning center can attract the vortex line. In
contrast to small defects, the pin-breaking force for large-size
defects is limited by the line tension of the vortices. With
increasing external force, the tips of the vortex line slide along
the surface of the insulating inclusion until they meet near the
equator and reconnect leading to the depinning of the vortex.
For the in-plane current in the anisotropic layered material, the
upper estimate for such a line-tension-limited force can be obtained
by considering the simple geometry of equally spaced pins aligned
along the $c$ axis and neglecting the interaction between vortex
tips at the pin surface. In this case, evaluating the external force
at which the tips meet, we obtain the following estimate
\begin{equation}
F_{p} \lesssim
(2\varepsilon_{0}/\gamma)\ln(b_z/\xi_{\min}),\label{PinBreakFrc}
\end{equation}
where $\gamma$ is the anisotropy factor, $\xi_{\min}=\max(\xi_c,s)$,
$\xi_c$ is the c-axis coherence length, and $s$ is the interlayer
period of a layered superconductor. This force only weakly depends
on the size and shape of the pinning center.

The interaction of the vortex line with a remote pin is long ranged,
due to the perturbation of the supercurrent flow around the vortex
by the pin,
\begin{equation}
U_{i}(r)\approx-\frac{\varepsilon_{0}V_{p}}{\pi(1-n_{y})r^{2}},
\text{ for } b\ll r\ll \lambda_{ab},\label{PinVortIntera}
\end{equation}
where $V_{p}= (4\pi/3) b_{z}b_{x}^{2}$ is the volume of the pinning
center and $n_{i}$ are ``depolarization factors'', which depend on
the parameter $\gamma b_{z}/b$. In particular, in the case $
b_{z}>b/\gamma$ which includes close-to-spherical inclusions,
\[
n_{z}  =\frac{1-\zeta^{2}}{\zeta^{3}}\left(  \tanh^{-1}\zeta-\zeta\right)  ,\
\text{with } \zeta=\sqrt{1-\frac{b^{2}}{\gamma^{2}b_{z}^{2}}}
\]
and $n_{x}=n_{y}=(1-n_{z})/2$.

Recently, it was demonstrated that magnetic force microscopy can be
effectively used not only for imaging but also for the manipulation
of individual vortices \cite{MFMManip}. This technique gives
principal possibility to measure the interaction between a vortex
line and an individual pinning center and extract the relevant
interaction parameters described in this section.

\section{Analytical estimates for the trapping of a vortex line \label{Sec-AnalEst}}

\subsection{General conditions for a static pinned line \label{SubSec-Cond}}

Consider a general vortex-line configuration trapped at the points
$(\mathbf{u}_{n},z_{n})$. For simplicity, we assume that the forces
from the pins are applied locally at the points $z=z_{n}$. In
between the trapped points, $z_{n}<z<z_{n+1}$, the displacement
obeys the following equation
\begin{equation}
\varepsilon_{1}\frac{\partial^{2}\mathbf{u}}{\partial
z^{2}}+f\mathbf{e} _{x}=0,\label{EqStatDispl}
\end{equation}
where $f$ is the driving force applied along the $x$ axis and
$\varepsilon _{1}=(\varepsilon_{0}/\gamma^{2})\mathcal{L}_{1}$ is
the line tension with $\mathcal{L}_{1}$ being the logarithmic factor
$\mathcal{L}_{1}=\ln(r_{\max }/r_{\min})$.\cite{NoteLogLength} The
displacement can be found as
\begin{equation}
\mathbf{u}(z)\!=\!\mathbf{u}_{n}\!+\!\left(
\mathbf{u}_{n+1}\!-\!\mathbf{u}_{n}\right)
\frac{z-z_{n}}{z_{n+1}\!-\!z_{n}}\!-\!\mathbf{e}_{x}
\frac{f(z\!-\!z_{n})(z\!-\!z_{n+1})}{2\varepsilon_{1}}
.\label{StatDispl}
\end{equation}
The force acting from the pin on the vortex line at $z=z_{n}$ is
given by
\[
\mathbf{F}_{n}\!=\!-\varepsilon_{1}\left[
\frac{\partial\mathbf{u}}{\partial
z}(z_{n}+0)-\frac{\partial\mathbf{u}}{\partial z}(z_{n}-0)\right]
\]
and can be evaluated as
\begin{equation}
\mathbf{F}_{n}\!=\!-\mathbf{e}_{x}\frac{f( z_{n+1}\!-\!z_{n-1})}{2}
-\varepsilon_{1}\left(
\frac{\mathbf{u}_{n+1}\!-\!\mathbf{u}_{n}}{z_{n+1}\!-\!z_{n}
}\!-\!\frac{\mathbf{u}_{n}\!-\!\mathbf{u}_{n-1}}{z_{n}\!-\!z_{n-1}}\right)
.\label{Fn}
\end{equation}
The stability condition for the trapped line is given by
\begin{equation}
F_{n}<F_{p}\text{ \ for all }n,\label{StabCond}
\end{equation}
while the critical state corresponds to the condition that at least
one local force reaches the pin-breaking force
\begin{equation}
\max_{n}\left(  F_{n}\right)  =F_{p}.\label{CritStateCond}
\end{equation}
One simple consequence of Eqs.\ (\ref{Fn}) and (\ref{StabCond}) is
that for ``behind'' sites, $u_{x,n}<u_{x,n-1},u_{x,n+1}$, the
line-tension force adds to the external force meaning that they, in
average, have  shorter trapping segments $z_{n+1}-z_{n}$,
$z_{n}-z_{n-1}$.

\subsection{Equilibrium trapping \label{SubSec-Equil}}

Consider the trapping of a single vortex line by strong-pinning centers with
concentration $n_{p}$ and pinning energy $U_{p}$, Eq.\ (\ref{PinEnergy}).
Assuming local equilibrium, the typical trapping length $L$ and transverse
displacement $u$ are determined by the energy-balance condition
\cite{BeekPRB02}
\[
\varepsilon_{1}\frac{u^{2}}{L}=U_{p}
\]
and by the condition that the average number of impurities in the
trapping volume should be of the order of one,
\[
n_{p}u^{2}L  =1.
\]
These equations give
\begin{equation}
L_{\mathrm{eq}}   =\sqrt{\frac{\varepsilon_{1}}{n_{p}U_{p}}},\ \
u_{\mathrm{eq}}^{2}
=\sqrt{\frac{U_{p}}{n_{p}\varepsilon_{1}}}.\label{L-u2-eq}
\end{equation}
Strictly speaking, the above conditions are obtained for zero current. Assuming
that the trapping length does not change much when current is applied, we
obtain an estimate for the critical current for the equilibrium trapping
\begin{equation}
\frac{\Phi_{0}}{c}j_{c,\mathrm{eq}}\approx\frac{F_{p}}{L_{e}}\approx
F_{p}\sqrt {\frac{n_{p}U_{p} }{\varepsilon_{1}}}.\label{EqCritCur}
\end{equation}
It is expected to increase with the pin density as $\sqrt {n_{p}}$.

\subsection{Dynamic trapping \label{SubSec-DynTrap}}

\begin{figure}[ptb]
\begin{center}
\includegraphics[width=3.2in]{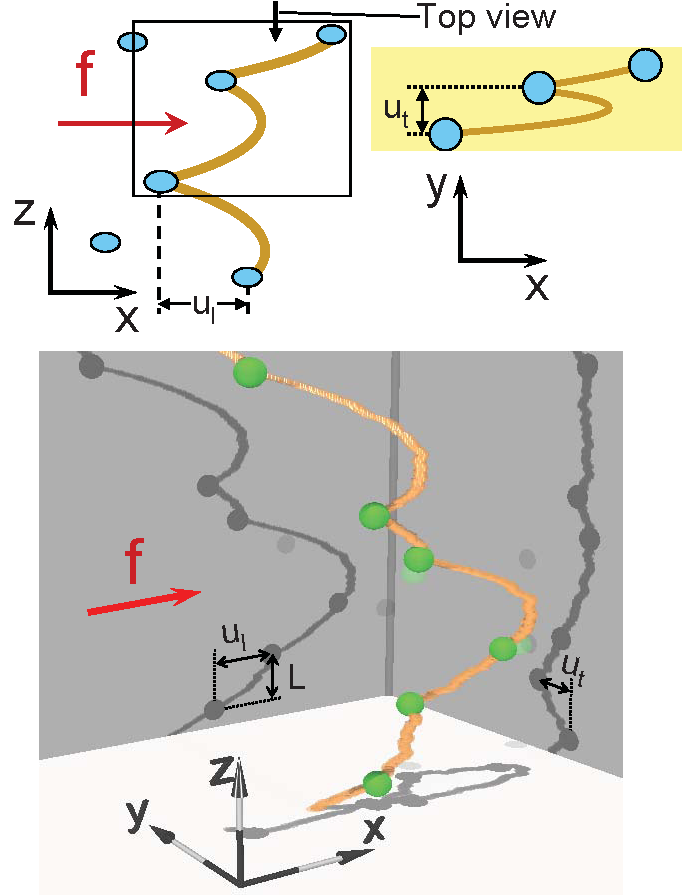}
\end{center}
\caption{\emph{Upper figures} illustrate a trapped vortex line in the
metastable regime (side and top views). The typical displacement in the
direction of the force $u_l$ is much larger than the typical displacement in
the perpendicular direction $u_t$. \emph{Lower figure} is a visualization of
the pinned line configuration obtained in simulations (only a short section of
the line is shown). Shadelike projections on the axis planes illustrate line
displacements in the different directions. Short scale line wiggling is due to
the thermal noise. The definitions of the trapping parameters $L_t$, $u_l$, and
$u_t$ are also illustrated.} \label{Fig-MetastableSingle}
\end{figure}

\begin{figure}[ptb]
\begin{center}
\includegraphics[width=2.4in]{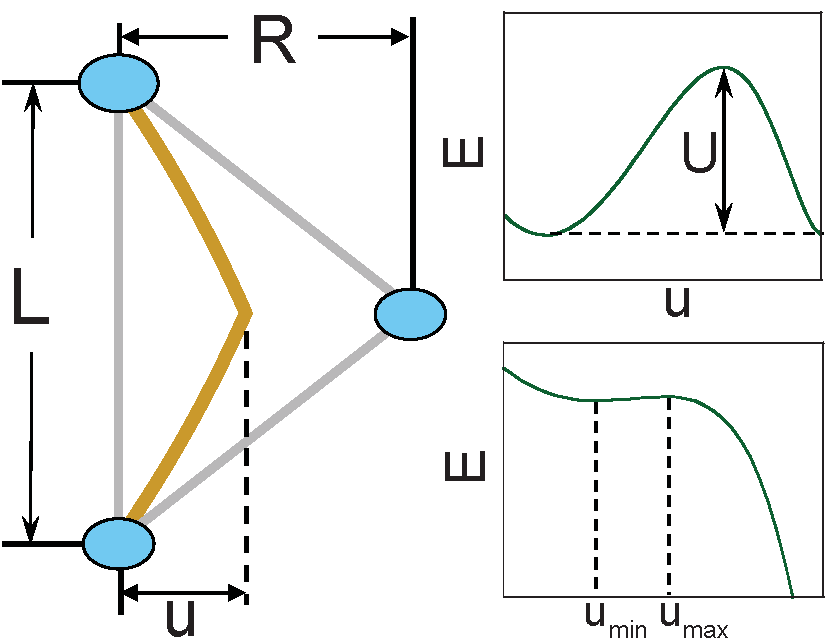}
\end{center}
\caption{A trapped segment interacting with pinning center. The upper right
plot illustrates the energy profile for the value of $R$ below which the
trapping becomes energetically favorable but separated by the energy barrier.
The lower right plot illustrates the energy profile near the trapping
instability point. } \label{Fig-SegmentInteraction}
\end{figure}

The equilibrium estimates for the trapping parameters (\ref{L-u2-eq}) are
definitely valid for the line configurations prepared by cooling at fixed field
and at zero transport current. However, it is clear that they cannot be applied
to the vortex lines in the critical state at low temperatures when moving lines
are trapped after the driving force drops below the critical value and
equilibration does not take place. The critical current in such a dynamic
regime has been estimated in Ref.\ \onlinecite{OvchinnIvlPRB91} for high fields
when the intervortex interactions are essential. These estimates can be
directly generalized to the trapping of individual vortex lines at small
fields.\cite{BlatterPRL04} When the vortex line moves close to the pinning
center, it may be trapped and the line remains trapped until the force acting
from the pin on the vortex line does not exceed the pin-breaking force. In this
regime two typical trapping distances, in the direction of motion, $u_{l}$, and
in the transverse direction, $u_{t}$, are very different and have very
different origins, see Fig.\ \ref{Fig-MetastableSingle}. These distances and
the trapped-segment length $L$ are connected by the geometric relation
\begin{equation}
n_{p}Lu_{l}u_{t}=1.\label{TrapMetaEq2}
\end{equation}
The longitudinal trapping distance, $u_{l}$, is determined by the
pin-breaking condition,
\begin{equation}
\varepsilon_{1}\frac{u_{l}}{L}=F_{p}.\label{TrapMetaEq1}
\end{equation}
This condition can be obtained from the $x$ component of Eq.\ (\ref{Fn})
assuming that the two terms on the right-hand side are of the order of the
pin-breaking force.

The transverse displacement between the pins, $u_{t}$, is determined by the
trapping events. The simplest assumption is that in most cases trapping occurs
when the lines directly collide with the pins \cite{BlatterPRL04} meaning that
$u_{t}\approx b$. This immediately gives estimates for other trapping
parameters,
\begin{equation}
u_{l}=\sqrt{\frac{F_{p}}{n_{p}\varepsilon_{1}b}},\ L=\sqrt{\frac
{\varepsilon_{1}}{n_{p}bF_{p}}}.\label{Trap-utb}
\end{equation}
This corresponds to the following result for the critical current
\begin{equation}
\frac{\Phi_{0}}{c}j_{c,\mathrm{tr}}=F_{p}^{3/2}\sqrt{\frac{n_{p}b}
{\varepsilon_{1}}}.\label{CrCurrMet_b}
\end{equation}
Note that in this situation the estimate is somewhat similar to the result for
the equilibrium case, Eq.\ (\ref{EqCritCur}), and has the same dependence on
the pin density, $\propto \sqrt{n_{p}}$. The physical assumptions behind the
two estimates, however, are completely different.

The above simple assumption, however, may underestimate $u_{t}$. Due to the
long-range pin-vortex interaction (\ref{PinVortIntera}) a pinning center may
capture the vortex line even without direct collisions. When the line passes
sufficiently close to the pinning center it may be trapped by this center due
to instability. To estimate the maximum transverse trapping distance $u_{t}$,
we consider the interaction energy of a segment of length $L$ with a pinning
center located at a distance $R\gg b$,\cite{OvchinnIvlPRB91} see Fig.
\ref{Fig-SegmentInteraction},
\[
E(u)=\varepsilon_{1}\frac{2u^{2}}{L}-\frac{A\varepsilon_{0}V_{p}}{(R-u)^{2}}
\]
with $A=1/[\pi(1-n_{y})]$. This gives the interaction force
\[
F(u)=\varepsilon_{1}\frac{4u}{L}-\frac{2A\varepsilon_{0}V_{p}}{(R-u)^{3}}
\]
Introducing the reduced variables
\[
x=\frac{u}{R},\
W=\frac{A\varepsilon_{0}LV_{p}}{2\varepsilon_{1}R^{4}},
\]
we rewrite the energy and force as
\begin{align*}
E &  =\varepsilon_{1}\frac{2R^{2}}{L}\left[
x^{2}-\frac{W}{(1-x)^{2}}\right]
,\\
F(x) &  =-\varepsilon_{1}\frac{4R}{L}\left(
x-\frac{W}{(1-x)^{3}}\right)  .
\end{align*}
The equilibrium points are determined by
\[
x(1-x)^{3}=W.
\]
The instability point corresponds to the value of $W$ when the
equilibrium points vanish which happens at
$W>W_{\max}=\max[x(1-x)^{3}]=3^{3}/4^{4}$. Therefore the condition
for the instability can be written as
\[
\frac{A\varepsilon_{0}LV_{p}}{2\varepsilon_{1}u_{t}^{4}}=\frac{3^{3}}{4^{4}},
\]
which determines the maximum trapping distance $u_{t}$ as
\begin{equation}
u_{t}=4\left(
\frac{A\varepsilon_{0}LV_{p}}{54\varepsilon_{1}}\right)
^{1/4}.\label{ut}
\end{equation}
The vortex line will be trapped by the pinning centers located
closer than this distance in the direction perpendicular to the
driving force. Note that the numerical coefficient in this equation
should not be taken too literally because it is only correct for the
simplest geometry illustrated in Fig.\ \ref{Fig-SegmentInteraction}.
Using this result, we find from Eqs.\ (\ref{TrapMetaEq2}) and
(\ref{TrapMetaEq1})
\begin{align}
L_{\mathrm{tr}} &  =\left[
\frac{\varepsilon_{1}^{5/4}}{n_{p}F_{p}\left(
\varepsilon_{0}V_{p}\right)  ^{1/4}}\right]  ^{4/9},\ \label{LtMeta}\\
u_{l} &  =\left[  \frac{F_{p}^{5/4}}{n_{p}\varepsilon_{1}\left(
\varepsilon_{0}V_{p}\right)  ^{1/4}}\right]  ^{4/9},\label{ulMeta}\\
u_{t} &  =\left[
\frac{\varepsilon_{0}^{2}V_{p}^{2}}{n_{p}\varepsilon
_{1}F_{p}}\right]  ^{1/9}.\label{utMeta}
\end{align}
This gives the following estimate for the critical current
\begin{equation}
\frac{\Phi_{0}}{c}j_{c,\mathrm{tr}}=\frac{n_{p}^{4/9}F_{p}^{13/9}\left(
\varepsilon_{0}V_{p}\right)  ^{1/9}}{\varepsilon_{1}^{5/9}}
\end{equation}
Comparing this result with the critical current for the equilibrium regime
(\ref{EqCritCur}), we can see that the two regimes are characterized by
somewhat different dependences on the parameters. However, dependences on the
pin density occur to be close, the exponent in the power law, $j_{c}\propto
n_{p}^{\alpha }$ in this dynamic-trapping regime is somewhat smaller than the
power $1/2$ for the equilibrium regime, $\alpha=4/9\approx0.444$.

\subsection{A typical pin-breaking force at finite temperatures \label{CritFrcT-est}}

At finite temperatures the line moves at all driving forces but for small
forces very slow motion occurs due to the rare thermally activated jumps (creep
regime). The creep and flow regimes are separated by the effective critical
force which can be evaluated using a velocity criterion. Such effective
critical force is criterion dependent and thus differs from the sharply defined
zero-temperature critical force of the depinning transition, but it has the
advantage that it can be directly compared with experimental estimates. We do
not consider creep in this paper and our purpose is to evaluate the suppression
of this critical force by the thermal noise. The main mechanism of thermal
suppression is reduction of the effective pin-breaking force. At finite
temperatures the trapped vortex segment has a finite lifetime on the pin even
if the pinning force $F$ is smaller than the maximum pin-breaking force. To
quantify this effect, we can introduce the temperature-dependent force
$\tilde{F}_{p}(T)<F_{p}$ at which the trapped segments are typically released
from pins.

To evaluate this force, we assume that the line motion in the crossover region
consists of segment jumps, meaning that the average line velocity can be
estimated as $ v\approx u_{l}/\tau, $ where $ \tau=\tau_{0}\exp\left[
U(F)/T\right] $ is the typical time during which the vortex segment remains
pinned and $U(F)$ is the typical energy barrier for detrapping. When the force
$F$ acting from the pin is only slightly smaller than the maximum pin-breaking
force, the barrier behaves as in the single-particle case,\cite{SinglePartNote}
$U(F)=a_{F}U_{p}(1-F/F_{p})^{3/2}$, where
$a_{F}=4\sqrt{2}F_{p}^{3/2}/(3U_{p}\sqrt{|F_{p}^{\prime\prime}|})$ and
$F_{p}^{\prime\prime}$ is the second derivative of the interaction force with
respect to the line displacement at the maximum-force point. The velocity
criterion for the effective critical force can be written as $\eta
v=C_{f}f_{c}$ where $\eta$ is the viscosity coefficient and $C_{f}\ll1$. At low
temperatures this gives us the following relation for $\tilde{F}_{p}$
\[
\frac{\tau_{0}}{\eta}\exp\left[
\frac{a_{F}U_{p}}{T}\left(1-\frac{\tilde{F}_{p}}{F_{p}}
\right)^{3/2}\right]  \approx\frac{u_{l}}{C_{f}f_{c}}.
\]
Using the estimate $f_{c}\approx\tilde{F}_{p}/L_{t}$, the geometric relation
(\ref{TrapMetaEq2}) and assuming for simplicity that $u_{t}\approx b$, we
obtain
\begin{equation}
\tilde{F}_{p}(T)\approx F_{p}\left[  1-\left(
\frac{T}{a_{F}U_{p}}\ln
\frac{\eta}{C_{f}\tau_{0}\tilde{F}_{p}n_{p}b}\right)  ^{2/3}\right].
\label{RenormFp}
\end{equation}
We expect that at low temperatures the effective critical force and trapping
parameters can be roughly evaluated using the simple replacement $F_{p}
\rightarrow\tilde{F}_{p}(T)$.

\section{Model for numerical simulations \label{Sec-NumModel}}

To develop a quantitative understanding of the strong pinning by the array of
inclusions, we numerically simulated motion of the vortex line described by the
dynamic equation
\begin{equation}
\eta\frac{\partial\mathbf{u}}{\partial
t}\!=\varepsilon_{1}\frac{\partial ^{2}\mathbf{u}}{\partial
z^{2}}\!+\!\sum_{j}\mathbf{F}(\mathbf{u}\!-\!\mathbf{R}
_{j})\delta(z\!-\!z_{j})+\mathbf{e}_xf+\mathbf{F}_{T}(z,t).\label{DynEq}
\end{equation}
Here $f$ is the driving force along the $x$ direction from the current,
$(\mathbf{R} _{j},z_j)$ are the random pin coordinates, $\mathbf{F}_{T}(z,t)$
is the Langevin thermal force
\[
\left\langle
F_{T,\alpha}(z,t)F_{T,\alpha^{\prime}}(z^{\prime},t^{\prime
})\right\rangle =2\eta
T\delta_{\alpha\alpha^{\prime}}\delta(t-t^{\prime
})\delta(z-z^{\prime}),
\]
and
\[
\mathbf{F}(\mathbf{u})=-\frac{\partial
U(\mathbf{u})}{\partial\mathbf{u}}
\]
is the interacting force with a strong pin. We model the interaction
potential by the function
\begin{equation}
U(\mathbf{u})=-\frac{U_{p}b^{2} G_{\mathrm{cut}}(u)}{u^{2}+b^{2}},
\label{ModelInteract}
\end{equation}
where the cutoff function $G_{\mathrm{cut}}(u)$ is introduced for numerical
convenience, $G_{\mathrm{cut}}(u)=(1-u^2/R_{\mathrm{cut}}^2)^2$ for
$u<R_{\mathrm{cut}}$ and $G_{\mathrm{cut}}(u)=0$ for $u>R_{\mathrm{cut}}$ with
$R_{\mathrm{cut}}\gg b$.   An important feature that was not taken into account
in modeling before is the long $1/u^{2}$ tail in the interaction potential.
However, this model does not describe the line-tension limited pin-breaking
force. In our model the pin-breaking force from an isolated pin is given by
$F_p=(3\sqrt{3}/8)U_p/b$ which is achieved at $u=b/\sqrt{3}$.  The model in its
original form has an unrealistic feature. For improbable configurations when
many pins are located at distances smaller than $b$, the vortex interaction
with such a cluster may increase without limit. This, of course, does not
happen in real superconductors. To bring our model somewhat closer to reality,
we renormalized the total pin-vortex interaction force as
$\mathbf{F}_{\mathrm{v-p}}(\mathbf{u})=\sum_{j}\mathbf{F}(\mathbf{u}\!-\!\mathbf{R}
_{j})$ as
\[
\mathbf{F}_{\mathrm{v-p}}\rightarrow \mathbf{F}_{\mathrm{v-p}}
\frac{\tanh(F_{\mathrm{v-p}}/F_{\mathrm{lim}})}{F_{\mathrm{v-p}}/F_{\mathrm{lim}}},
\]
so that the maximum force cannot exceed $F_{\mathrm{lim}}$. This modification
has only a minor influence on the interaction of the vortex line with an
isolated pin.

For numerical implementation of the model,  we use the reduced
variables
\begin{align*}
\mathbf{u}  &  =b\tilde{\mathbf{u}},\
z=\frac{\varepsilon_{1}b^{2}}{U_{p}
}\tilde{z},\ t=\frac{\eta\varepsilon_{1}b^{4}}{U_{p}^{2}}\tilde{t},\\
\mathbf{f}  &
=\frac{U_{p}^{2}}{\varepsilon_{1}b^{3}}\tilde{\mathbf{f} },\
\tilde{\mathbf{F}}(\tilde{\mathbf{u}})=\frac{\partial}{\partial
\tilde{\mathbf{u}}}\frac{G_{\mathrm{cut}}(\tilde{u})}{\tilde{u}^{2}+1},
\end{align*}
in which the equation takes the simpler form
\begin{equation}
\frac{\partial\tilde{\mathbf{u}}}{\partial\tilde{t}}=\frac{\partial^{2}
\tilde{\mathbf{u}}}{\partial\tilde{z}^{2}}+\sum_{j}\tilde{\mathbf{F}}
(\tilde{\mathbf{u}}-\tilde{\mathbf{R}}_{j})\delta(\tilde{z}-\tilde{z}
_{j})+\mathbf{e}_x\tilde{f}+\tilde{\mathbf{F}}_{T}(z,t)
\label{RedEq}
\end{equation}
with
\begin{align*}
\left\langle
\tilde{F}_{T,\alpha}(\tilde{z},\tilde{t})\tilde{F}_{T,\alpha
^{\prime}}(\tilde{z}^{\prime},\tilde{t}^{\prime})\right\rangle  &
=2\tilde
{T}\delta_{\alpha\alpha^{\prime}}\delta(\tilde{t}-\tilde{t}^{\prime}
)\delta(\tilde{z}-\tilde{z}^{\prime})\\
\tilde{T}  &  =T/U_{p}
\end{align*}
In this dimensionless form, the equation depends only on the reduced
temperature and reduced pin density $\tilde{n}_{p}=(\varepsilon_{1}b^{4}/U_{p}
)n_{p}$. The condition of the strong pinning regime is $n_{p}<U_{p}/\varepsilon
_{1}b^{4}$ corresponding to $\tilde{n}_{p}<1$. For the typical parameters
$\gamma=5$, $b_{z}=b=10$nm, and $n_{p}=(100 $nm$)^{-3}=10^{15}$cm$^{-3}$, the
reduced pin density can be estimated as $\tilde{n}_{p}\approx
n_{p}b^{4}/(b_{z}\gamma^{2}) \approx 10^{-4}$. In simulations we mostly used
$\tilde{F}_{\mathrm{lim}}=1$. For an isolated pin this gives the pin-breaking
force $\tilde{F}_{p}=\tanh(3\sqrt{3}/8)\approx 0.57$. We also used
$\tilde{R}_{\mathrm{cut}}=50$ in the cutoff function
$G_{\mathrm{cut}}(\tilde{u})$. We consider systems of size $L_x \times L_y
\times L_z$, where $L_z$ is the vortex-line length, $L_x$ is the size in the
direction of line motion, and $L_y$ the size in the transverse direction.

One of our tasks is to compute the dependence of the critical force on the
density of the pinning centers. The calculation of the steady-state critical
force is not as trivial problem as it may appear. For a finite-size system
there is always a metastable configuration giving the maximum depinning force.
Assuming that such a configuration is always reachable from any initial
condition at long times, this would be the critical force we seek. In a large
system this configuration is, however, determined by a rare, nontypical
configuration of pins. As a consequence, the maximum depinning force slowly
grows with the increasing the system sizes $L_x$ and $L_y$, meaning that it is
not a self-averaging parameter for a fixed line length $L_z$. To avoid this, a
carefully chosen (anisotropic) thermodynamic limit in all directions was
proposed for $d+1$ dimensions~\cite{RossoBolechPRL04}. The maximum depinning
force of a very large system is also very difficult to compute. Simulations of
Eq.\ (\ref{RedEq}) at zero temperature are not very suitable for this purpose,
because when the external force is close to the critical value, the vortex line
traps forever in the first metastable state it finds. It is not clear how
representative this state is and how close the corresponding critical force is
to the maximum value. In addition, neither the maximum critical force nor
depinning forces for a few accidental metastable states are very interesting
quantities from a practical point of view. The maximum critical force is
essentially a property of an isolated vortex line. A more interesting quantity
is the typical pinning force for the finite density of the vortex lines.
Indeed, even at low densities when vortex-vortex interactions can be neglected,
we expect a typical pinning force rather than an extreme non-self-averaging
force value to determine the observable critical current.

\begin{figure}[ptb]
\begin{center}
\includegraphics[width=1.64in]{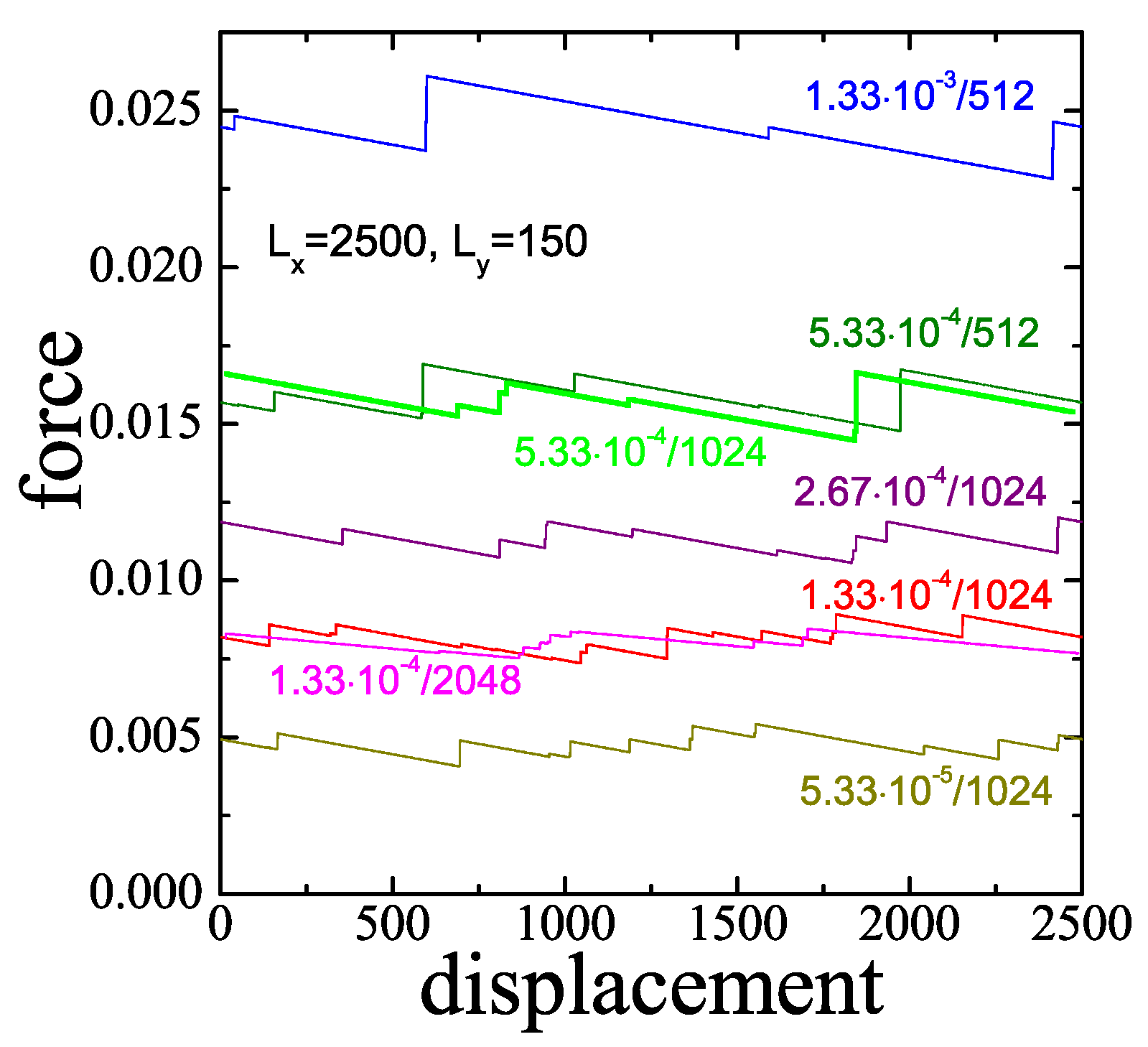}\includegraphics[width=1.75in]{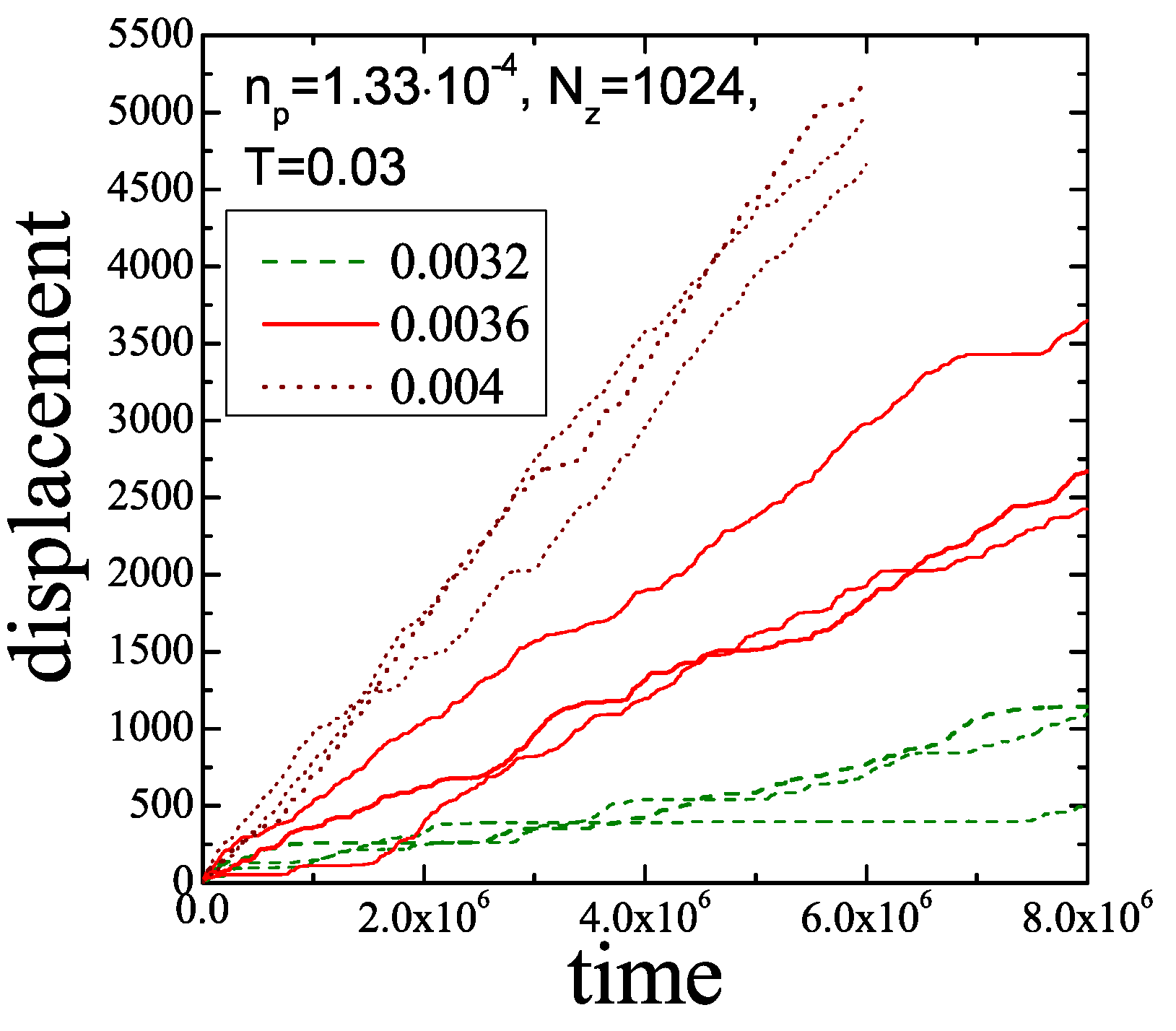}
\end{center}
\caption{\emph{Left:} Representative dependences of the force acting on the
line versus its center of mass location obtained using the fixed-velocity
simulations for different pin densities $n_{p}$ and line lengths $N_z$  (the
curves are marked by $n_p/N_z$). The vertical segments correspond to trapped
states. The typical critical forces, $f_c$, are obtained by averaging over the
local maxima of these curves. \emph{Right:} Examples of the displacement-time
dependences used to evaluate average velocities at fixed temperature for the
parameters shown in the plot and for different forces. Each force is
represented by three curves corresponding to different realizations of the
random potential. The line motion becomes more and more uneven with decreasing
force.} \label{Fig-Displ-Force-Time}
\end{figure}

To evaluate a typical pinning force at zero temperature, instead of the
fixed-force approach, we employ fixed-velocity simulations using approach
suggested in Ref.\ \onlinecite{MovingParabola}. Namely, we replaced the fixed
external force $\tilde{f}$ in Eq.\ (\ref{RedEq}) with a slowly moving parabolic
potential,
\begin{equation}
f\rightarrow K\left [W(t)-u_x(z,t)\right ],\
W(t)=W_0+Vt\label{MovingParab}
\end{equation}
Such a potential forces the vortex line to move with the average velocity $V$.
Every time the line finds a metastable pinned state and stops, the dragging
force starts to increase with time until it exceeds the critical force for this
state and the line resumes motion. This trick allows us to explore many
metastable states and to avoid the extreme value statistics of the
sample-dependent critical force. The typical critical force is then evaluated
as the average force acting on the vortex line in the critical configurations,
\begin{equation}
f_c= \langle\langle K\left [W(t)-u_x(z,t)\right
]\rangle_z\rangle_t\label{fcMovingParab}
\end{equation}
where $\langle\ldots\rangle_t$ implies averaging over the local maxima of the
instantaneous force. The spring constant $K$ and the drag velocity $V$ have to
be taken sufficiently small so that they do not influence the calculation of
the critical force. We typically use $K\sim 10^{-5}\!-\!10^{-6}$ and
$V=0.001\!-\!            0.002$. The spring constants satisfy $K \sim L_z^{-2}$
in each case, assuring a proper thermodynamic limit for the critical force and
associated critical configuration~\cite{MovingParabola}, and the velocities are
small enough to assure a quasistatic stick-slip motion. Figure
\ref{Fig-Displ-Force-Time}(left) illustrates the typical dependences of the
force acting on the line on the displacement of its center of mass for
different pin densities $n_{p}$ and line lengths $N_z$.

We also explore the velocity-force dependences at finite temperatures using the
direct fixed-force simulations described by Eq.\ (\ref{RedEq}). Even though the
term ``\emph{critical force}'' is widely used in the experimental community, at
finite temperatures the concept of the critical force does not have an exact
meaning because the velocity is finite at all forces due to the thermal creep.
Nevertheless, one can still introduce the characteristic force describing
crossover between the flux-flow and flux-creep regimes using some
average-velocity criterion. At low temperatures near such critical force the
line motion becomes very uneven, see Fig.\ \ref{Fig-Displ-Force-Time}(right).
It spends considerable time in metastable traps waiting for a strong
fluctuation which allows it to continue motion. Such line motion is illustrated
by animation.\cite{Animation} As a consequence, a proper averaging over such
events requires huge simulation times and/or averaging over many realizations
of the random potential. In addition to the critical force, we explore
statistical parameters of trapping which allow us to understand better the
pinning mechanism. We evaluated the average length of a trapped segment $L_{t}$
and typical displacements along the motion direction $u_{l}$ and in the
perpendicular direction $u_{t}$. We studied the distribution of these
parameters and their evolution with temperature and force.  We also studied the
long-range wandering of the line in the direction of motion and in the
transverse direction.

For a numerical solution the reduced equation (\ref{RedEq}) has to be
discretized both in time and in the $z$ coordinate. We typically used
$dt=0.05\!-\!0.1$ and $dz=1$ for the discretization steps. To study finite-size
effects, the equation was solved for different numbers of $z$-axis slices, $N_z
\equiv L_z/dz$, ranging from 512 to 2048.

\section{Numerical results \label{Sec-NumResults}}

\subsection{Behavior at zero temperature \label{SubSec-zeroT}}

\begin{figure}[ht]
\begin{center}
\includegraphics[width=3.0in]{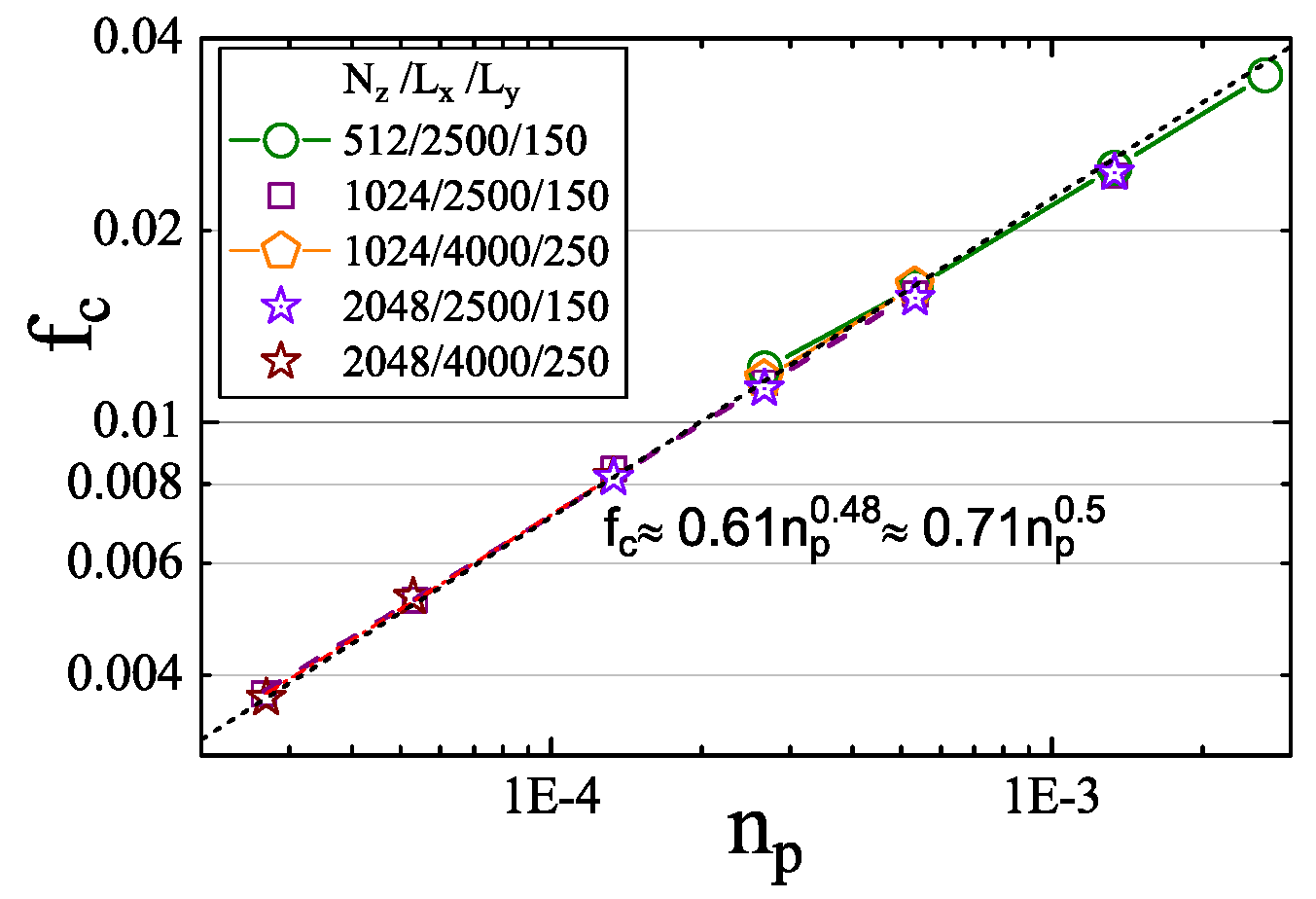}
\end{center}
\caption{The dependence of the critical force on the pin density. The plot
contains data obtained for different system sizes, and the legend shows
$N_z/L_x/L_y$. The finite-size effects in the critical force are weak.}
\label{Fig-fc-np}
\end{figure}
\begin{figure}[ht]
\begin{center}
\includegraphics[width=2.5in]{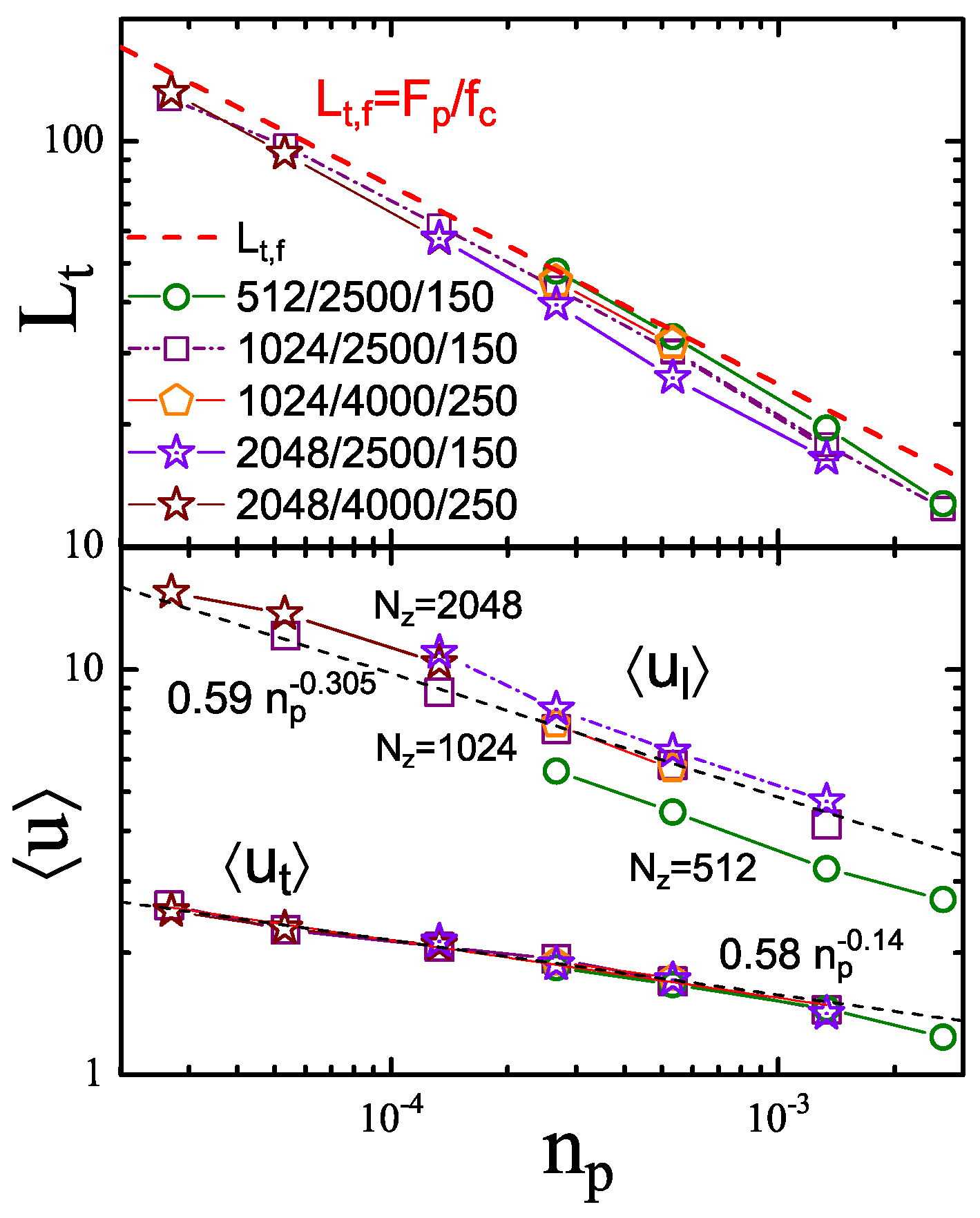}
\end{center}
\caption{Summary of the pin-density dependences of the trapping parameters for
different system sizes. As in the previous plot, the legend shows
$N_z/L_x/L_y$. \emph{Upper plot:} The dependences of the average
trapped-segment length $L_t$. For comparison, we show the expected trapping
length extracted from the critical force. \emph{Lower plot:} The pin-density
dependences of the average pin-to-pin displacements in the direction of the
force (longitudinal) and in the perpendicular direction (transverse). The
longitudinal displacement, $\langle u_l\rangle$, has a considerable finite-size
effect with respect to the line length $N_z$.} \label{Fig-Trap-param-np}
\end{figure}
\begin{figure}[ht]
\begin{center}
\includegraphics[width=2.5in]{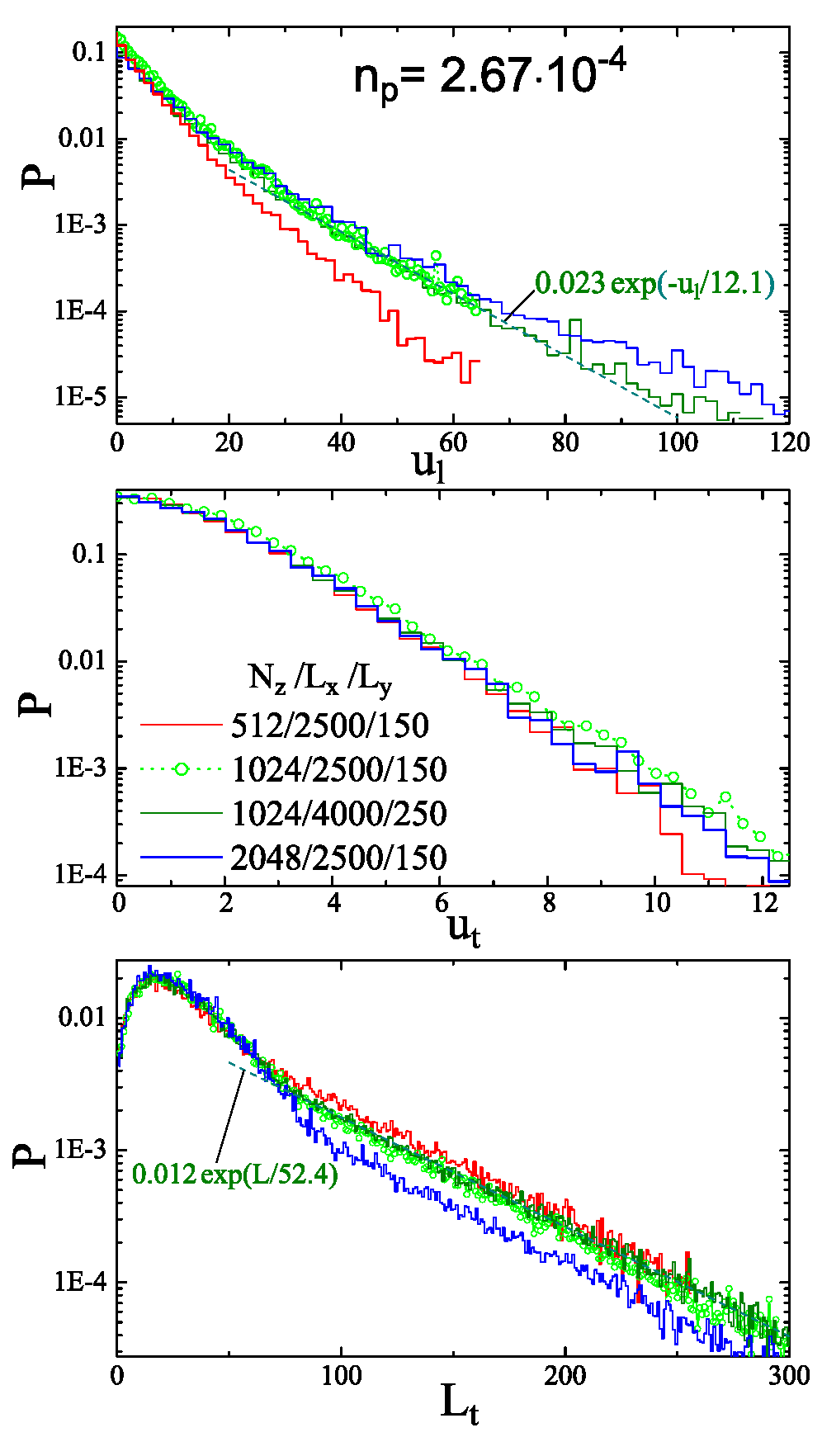}
\end{center}
\caption{Examples of distribution functions of trapping parameters
at one pin density and for different system sizes. }
\label{Fig-Distrib}
\end{figure}

We systematically studied the behavior of the critical force and the properties
of trapped line configurations within a  pin-density range spanning two orders
of magnitude, from $2.7\cdot 10^{-5}$ to $2.7\cdot 10^{-3}$. Figure
\ref{Fig-fc-np} presents the dependences of the critical force $f_c$ on the pin
density $n_p$ for different systems sizes $N_z$. We found that $f_c$ increases
with $n_p$ according to the power law $f_c=f_0 n_p^\alpha$. The fit gives for
the power index a value slightly smaller than 1/2, $\alpha\approx 0.48$ and the
coefficient $f_0\approx 0.61$. In fact, the square-root dependence $f_c\approx
0.71 \sqrt{n_p}$ also provides a reasonable description of the data. The power
is, however, clearly larger than the value $0.44$ suggested by the
dynamic-trapping estimates in the case when trapping occurs due to
instabilities. For the used line lengths $N_z \geq 512$ a noticeable
finite-size effect becomes visible only for small pin densities $n_p< 3\cdot
10^{-4}$.

To understand the statistical properties of trapped configurations, we plot in
Fig.\ \ref{Fig-Trap-param-np} the pin-density dependences of the average
trapping parameters defined in Fig.\ \ref{Fig-MetastableSingle}, the trapping
length $L_t$, and the pin-to-pin displacements along the direction of force,
$u_l$, and in the transverse direction, $u_t$. In the plot $L_t(n_p)$ we show
the expected value of the typical trapped segment extracted from the value of
the critical force $L_{t,f}=F_p/f_c$. We can see that the real trapping
segments extracted from the configurations indeed closely follow  the expected
values. We also see that the average trapping segments are systematically
smaller than the segments which determine the critical force. This is a natural
behavior because one can expect that the critical force is determined by ``weak
spots'' where the trapping segments are longer than average along the line. The
difference, however, is not very significant. The trapping length shows a weak
but unexpected size effect, it slightly \emph{decreases} with increasing total
line length. We found that the product $n_pL_tu_lu_t$, which on general ground
is expected to be of the order of unity is, in fact, slowly decreases with
$n_p$ from $\sim 0.18$ to $\sim 0.14$.

From the plots of the pin-to-pin displacements, we can see that the pinned
configurations are strongly anisotropic, the average displacement along the
direction of the driving force $u_l$ significantly exceeds the displacement in
the transverse direction $u_t$. This is consistent with the dynamic-trapping
picture described in Sec.\ \ref{SubSec-DynTrap}. The difference grows with
decreasing pin density $n_p$. In addition, the longitudinal displacement shows
a significant size effect, it grows with an increasing system size $N_z$.
Further analysis shows that this is an indication of growing local stresses
with increasing line length and suggests the destruction of the vortex lines by
the pinning potential in the critical state for sufficiently large systems.
This behavior is not anticipated by the simple estimates. It is interesting to
note, however, that this size effect in $u_l$ does not lead to a significant
size dependence of the critical force. On the other hand, the average
transverse displacement $u_t$ does not show any size effect. It slowly grows
with decreasing $n_p$ from 1.24 at $n_p=0.0027$ to 2.54 at $n_p=2.7\cdot
10^{-5}$. As $u_t$ remains comparable with the defect size, the regime in which
the transverse trapping is determined by the long $1/r^2$ tail of the
pin-vortex interaction is not quite realized. This explains why the power index
in the force$-$pin-density dependence is larger than suggested by the
metastable-regime estimates which assume $u_t\gg 1$. To obtain further insight
into the properties of trapped lines we show in Fig.\ \ref{Fig-Distrib}
examples of the distribution functions of trapping parameters for A fixed
density $n_p=2.67\cdot 10^{-4}$ and different system sizes. One can see that
these distributions are characterized by long exponential tails. There is a
noticeable probability to find segments with very large $u_l$ and $L_t$. Note
that the tails have opposite size effects for these parameters, the probability
to find a large $u_l$ increases with $N_z$ while the probability to find a
large $L_t$ \emph{decreases} with $N_z$. The last trend is opposite to naive
expectations.

\begin{figure*}[ptb]
\begin{center}
\includegraphics[width=5in]{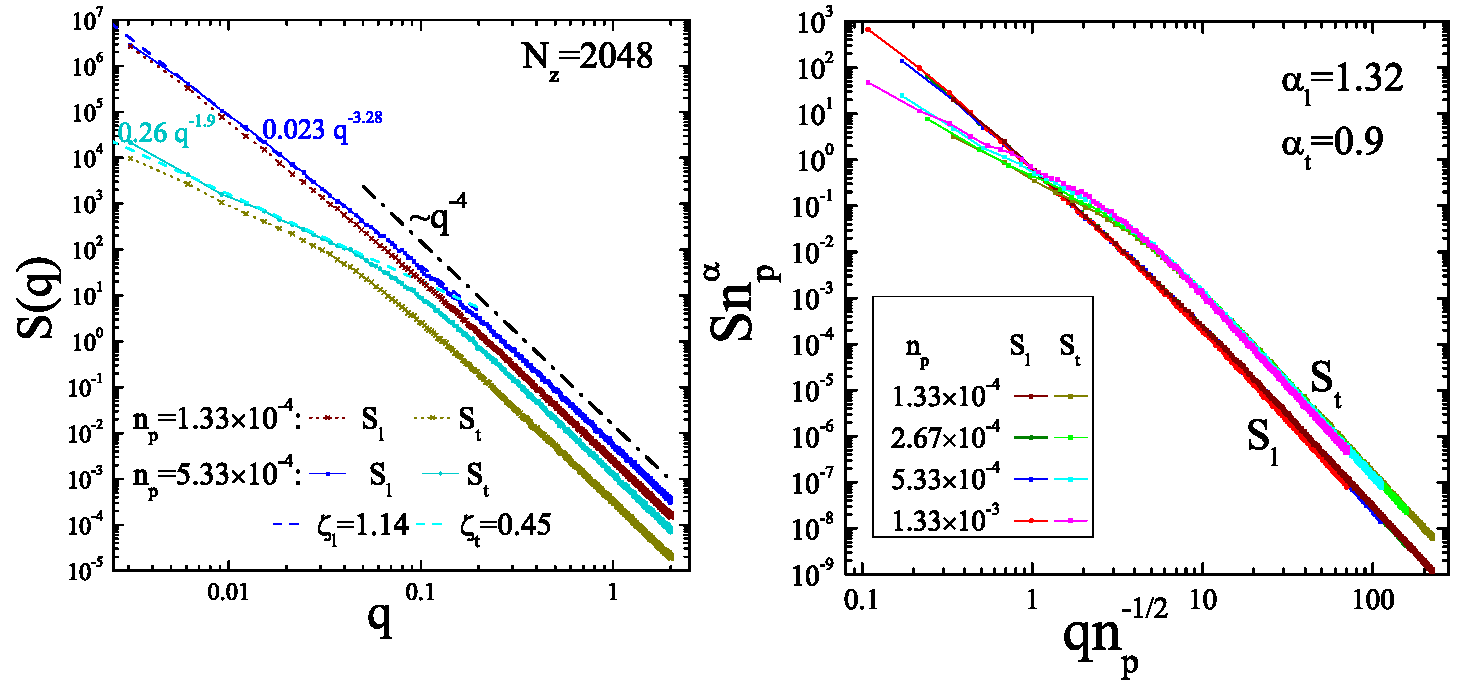}
\end{center}
\caption{\emph{Left:} Transverse and longitudinal structure factors of the line
for $N_z=2048$ and two densities of pins. Two regions of power-law $q$
dependences are clearly observed for both components. \emph{Right:} Scaled
structure factors for different pin densities.} \label{Fig-SqT0}
\end{figure*}
To characterize the long-range displacements of the line, we studied
behavior of the structure factors, the Fourier transforms of the
displacement correlation functions,
\begin{equation}
S_{l,t}(q)=\frac{1}{N_z}\left \langle \left
|\sum\limits_{z=1}^{N_z}u_{x,y}(z)\exp(-iqz) \right |^2\right
\rangle .\label{DefStruc}
\end{equation}
Examples of these quantities are presented in Fig.\ \ref{Fig-SqT0} (left) for
two pin densities. Similar to local quantities, the long-range displacements
are also strongly anisotropic. Such anisotropic scaling of the displacements is
a general property of the driven lines in the critical regime independent of
the pinning mechanism.\cite{ErtasKardarPRB} For both components we clearly
observe two regions of $q$ characterized by different power-law dependences
$S_{l,t}\propto q^{1+2\zeta_{l,t}}$. For the smallest $q$'s we found the
roughness exponents $\zeta_{l}\approx 1.14$ and $\zeta_{l}\approx 0.45$. The
value of $\zeta_t<1$ for the transverse direction corresponds to the line
displacements increasing as $\langle [u_y(z)-u_y(0)]^2\rangle\propto
z^{2\zeta_t}$ at large $z$. On the other hand, the value of $\zeta_l>1$ found
for the longitudinal displacements implies that the assumed elastic
approximation is not self-consistent in the thermodynamic limit, the average
local stress $\langle(du_x/dz)^2\rangle$ increases with the line length $L_z$
as $L_z^{2(\zeta_l-1)}$.  In this case the longitudinal line displacements grow
quadratically $\langle [u_y(z)-u_y(0)]^2\rangle= C_l z^{2}$ with the
coefficient increasing with the line length as $C_l\propto
L_z^{2(\zeta_l-1)}$.\cite{LeschhornTangComment93} This provides a natural
explanation for the strong size dependence of the longitudinal pin-to-pin
displacement $u_l$ in the lower plot of Fig.\ \ref{Fig-Trap-param-np}. The
negative size effects for the trapping lengths in the upper plot of the same
figure can also be understood. Growing local stress with increasing $N_z$
forces the line to travel longer distances in the longitudinal direction which
increases the probability of finding a pin separated by a smaller distance in
the $z$ direction.

The found exponents are slightly different from the values $\zeta_l=1$ and
$\zeta_t=0.5$ obtained in Ref.\ \onlinecite{ErtasKardarPRB} from the
approximate functional renormalization group calculations. However, a similar
situation was found for elastic lines in a plane where the numerically computed
index $\zeta=1.25$ \ \cite{RossoPRE03,2Dline} also exceed the predicted value
$\zeta=1$ \ \cite{2DzetaTheory}. We see that the transverse displacements
somewhat reduce the exponent value in the 3D case. Since the small-$q$
exponents are expected to be universal, i.e., independent of the pinning
mechanism, our results suggest that the elastic description will also break
down for the weak pinning case. Although we have considered the linear
approximation for the elastic forces, this conclusion is expected to hold for
the full non-quadratic energy of the deformed vortex line.\cite{RossoPRE03}

At larger $q$'s we observe the regime where both components behave as $S_{l,t}=
A_{t,l}q^{-4}$ giving the short-scale indices $\zeta_l=\zeta_t=3/2$. This
corresponds to displacements induced by a short-range-correlated random force,
and such behavior is actually similar to the static Larkin regime for weak
pinning. However, the random forces in our case clearly have a very different
origin.

The crossover between different regimes occurs at a wave vector that scales
approximately as $\sqrt{n_p}$. This allows us to approximately collapse the
structure factors at different pin densities into a single curve using the
scaling $S_{l,t}(q)=n_p^{-\alpha_{l,t}} G_{l,t}(qn_p^{-1/2})$. We found
$\alpha_l\approx 1.32$ and $\alpha_t\approx 0.9$. These scaled dependencies are
shown on the right-hand side of Fig.\ \ref{Fig-SqT0}. Scaling works better for
the longitudinal structure factor.  This scaling is consistent with the
identification of the trapping parameter $L_t(n_p)$ as a geometric crossover
length at $q L_t\sim 1$, between a short-distance roughness regime with
exponents $\zeta_l=\zeta_t=3/2$ and a large-distance universal regime with
exponents $\zeta_t \approx 0.45$ and $\zeta_l \approx 1.14$. This behavior is
again very similar to the crossover between the Larkin regime and the
random-manifold regime established for the case of weak collective pinning even
though disorder is not weak in the present case.

We can also see from Fig. \ref{Fig-SqT0}(left) that both structure factors
increase with the pin density for all wave vectors. This means that, in
contrast to elemental pin-to-pin displacements $u_{l,t}$ plotted in Fig.
\ref{Fig-Trap-param-np}, the line displacements at fixed $z$ \emph{grow} with
increasing pin density. This behavior can be easily understood. The
displacements at small distances are determined not only by the behavior of
$u_{l,t}$ but also by the behavior of the trapping length $L_{t}$. All these
parameters decrease with increasing pin density, meaning that smaller
displacements occur on a smaller length scale. The net increase of the line
displacements at fixed $z$ coordinate with increasing $n_p$ is a consequence of
a faster $L_{t}(n_p)$ decrease than that of $u_{l,t}(n_p)$,  as can be seen
from Fig. \ref{Fig-Trap-param-np}.

\subsection{Dynamics at finite temperatures \label{SubSec-finiteT}}

\begin{figure*}[ptb]
\begin{center}
\includegraphics[width=2.64in]{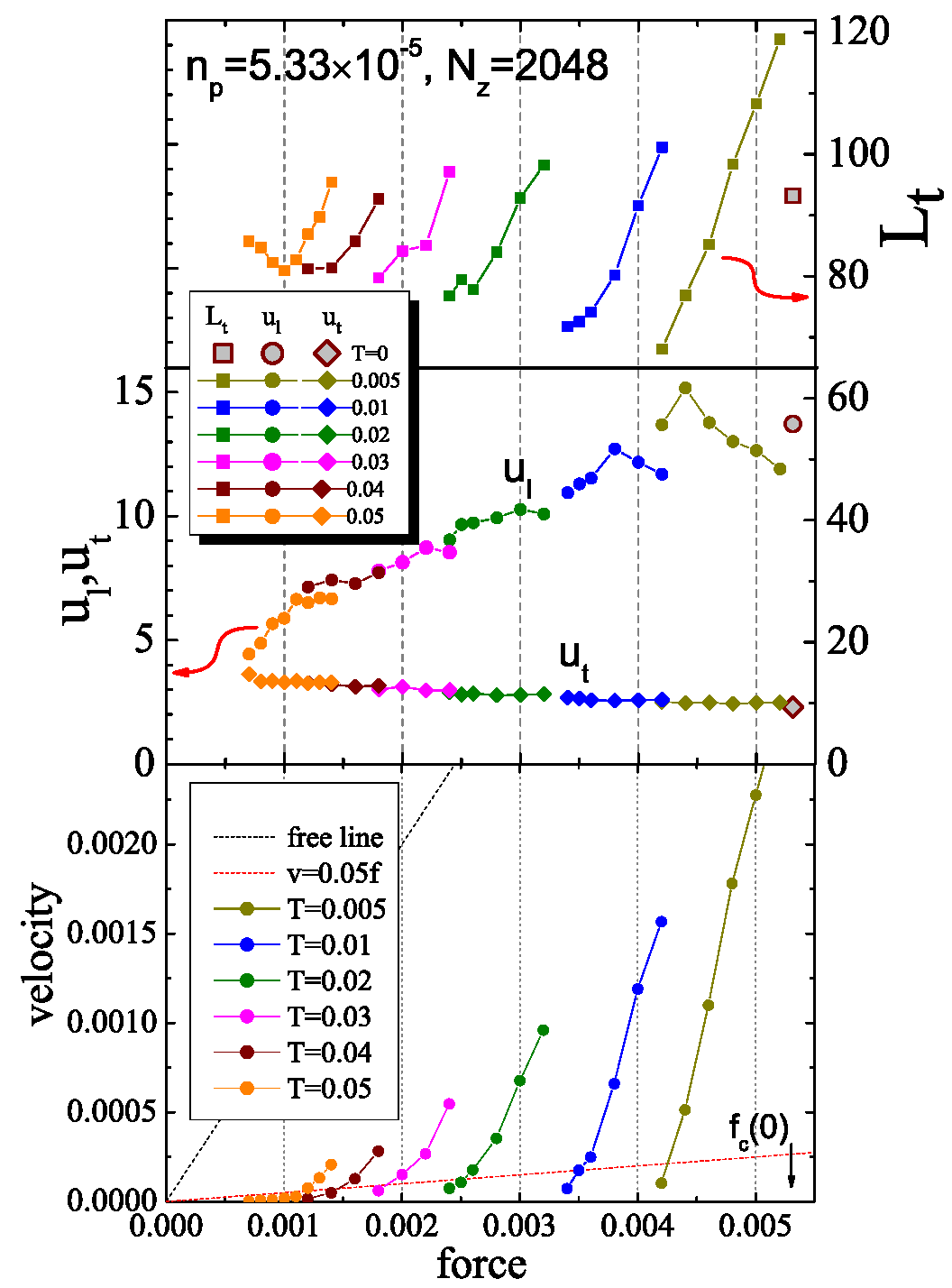}
\includegraphics[width=2.7in]{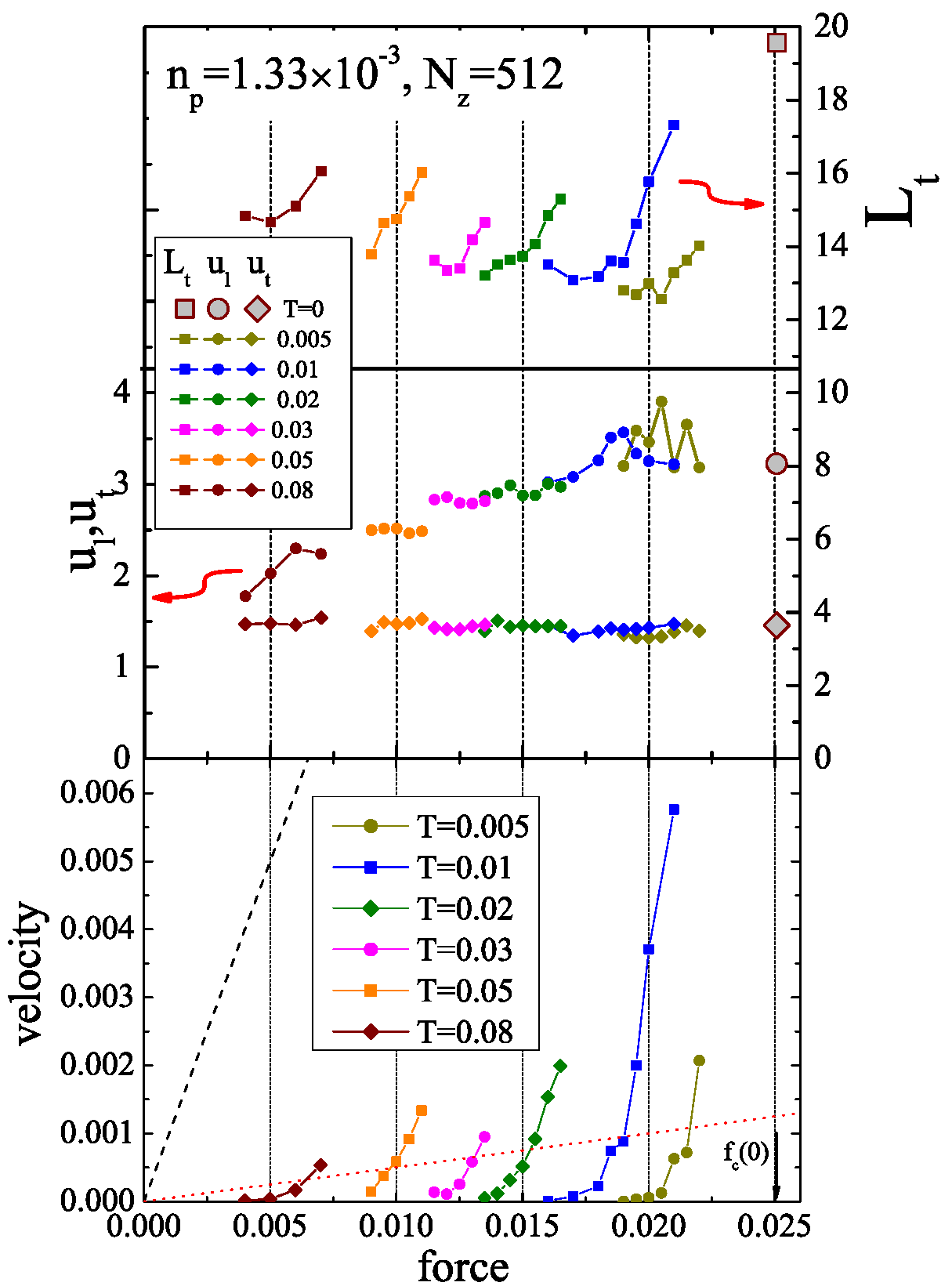}
\end{center}
\caption{Temperature evolution of the velocity-force dependences (bottom
plots), lengths of trapped segments (top plots, right axes), and pin-to-pin
displacements (middle plots) for two very different pin densities,
$n_p=5.33\times10^{-5}$ (left) and $1.33\times10^{-3}$ (right) in the regions
corresponding to crossover between flow and creep.}
\label{Fig-IVTrapParamTN20Nz2048}
\end{figure*}
\begin{figure*}[ptb]
\begin{center}
\includegraphics[width=6.8in]{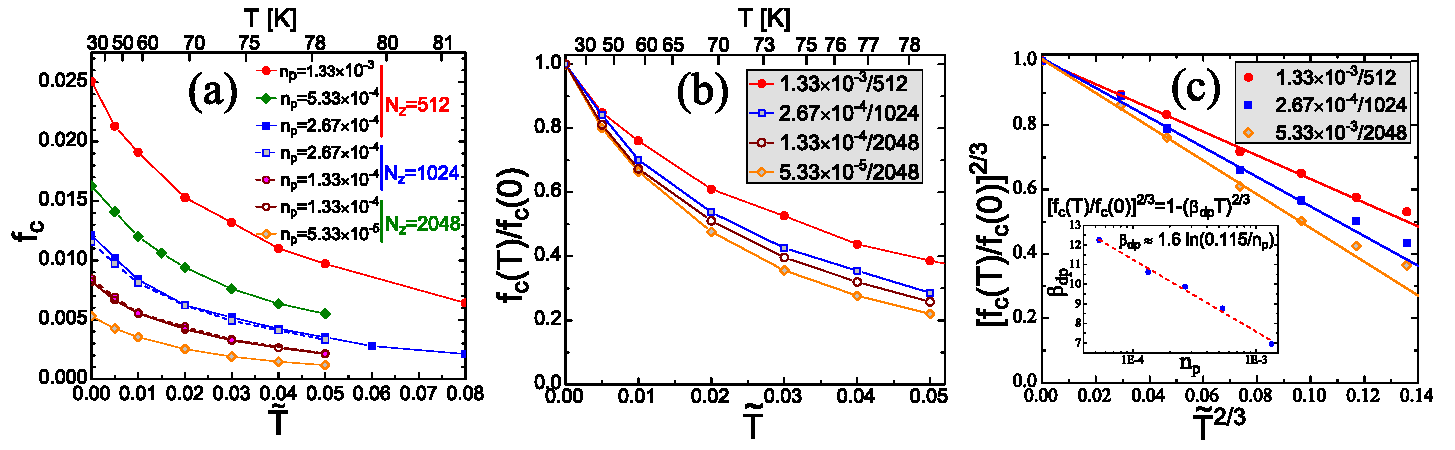}
\end{center}
\caption{(a)  Temperature dependences of the effective critical forces for
different pin densities and line lengths. (b) Relative suppression of effective
critical forces by thermal fluctuations for selected pin densities/line
lengths. (c) Plots of $[f_c(T)/f_c(0)]^{2/3}$ vs $T^{2/3}$ showing
approximately linear dependences in agreement with Eq.\ (\ref{RenormFp}). The
inset shows the pin-density dependence of the coefficient $\beta_{dp}$ with the
logarithmic fit. Legends in the plots (b) and (c) imply $n_p/N_z$. The top axes
in the plots (a) and (b) provide the approximate real temperature scales
computed for spherical particles with radius $b=5$ nm assuming
$\lambda_{ab}=(140$ nm$)/\sqrt{1-(T/T_c)^2}$ with $T_c=90$ K. }
\label{Fig-fc-T}
\end{figure*}
\begin{figure}[ptb]
\begin{center}
\includegraphics[width=3.5in]{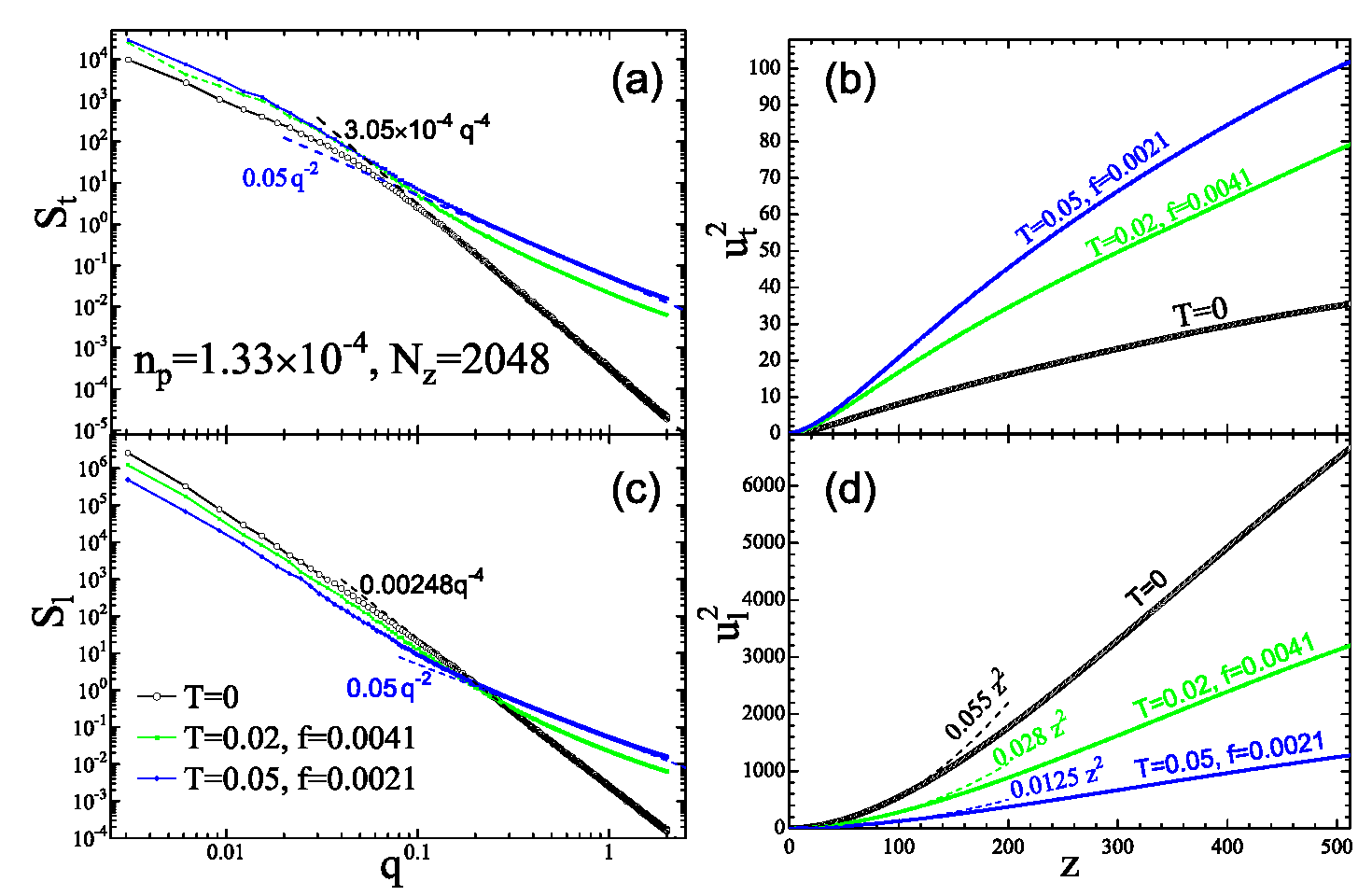}
\end{center}
\caption{ Typical evolution of the structure factors  (plots (a) and
(c)) and line displacements (plots (b) and (d)) with increasing
temperature for one pin density.   Thermal fluctuations in the
critical regime reduce the line displacements in the direction of
drive and increase the line displacements in the transverse
direction. } \label{Fig-Sl-ul2-T}
\end{figure}

In this section we consider the influence of thermal noise on the dynamic
response and configurations of vortex lines interacting with strong pins.
Figure \ref{Fig-IVTrapParamTN20Nz2048} presents the temperature evolution of
the velocity-force dependences and trapping parameters in the critical region
for two very different pin densities, $n_p=5.33\times 10^{-5}$ and
$n_p=1.33\times 10^{-3}$. We remind that we use the pinning energy of a single
pin as the temperature unit. At finite temperatures the critical force does not
have an exact meaning, because the line velocity is finite at all forces due to
the thermal creep. Nevertheless, we can introduce a typical force corresponding
to the crossover between the flow and creep regimes, similar to the voltage
criterion widely used in experiment. We use the criterion $v=0.05f$ for this
force. The first important observation is that, independently of criterion, the
apparent critical force is quite strongly suppressed by the thermal noise. For
example, as we can see in Fig.\ \ref{Fig-IVTrapParamTN20Nz2048}, for the small
pin density $n_p=5.33\times 10^{-5}$ at the temperature only 5\% of the pinning
energy, the apparent critical force is already suppressed about fourfold. As
one can see from the trapping-parameter plots, the longitudinal pin-to-pin
displacement $u_l$ in the critical region very rapidly decreases with the
temperature while the transverse displacement $u_t$ slightly increases with the
temperature. As a consequence, at some temperature  they become of the same
order. The longitudinal displacement typically has a nonmonotonic dependence on
the driving force and reaches a maximum at some force in the critical region
(the maximum-stress force). The trapping length decreases with decreasing line
velocity and its value at $v=0.05f$ slightly increases with the temperature.

Figure \ref{Fig-fc-T}(a) shows the temperature dependences of the apparent
critical forces for a wide range of pin densities. We can see that these
dependences are quite similar. However, plots of the relative critical forces
in Fig.\ \ref{Fig-fc-T}(b) clearly show that thermal suppression weakens with
increasing pin density. This is consistent with the estimate for the
temperature renormalization of the effective pin-breaking force, Eq.\
(\ref{RenormFp}), due to the $n_p$ dependence under the logarithm. For
illustration, we also present  the real-temperature scales on the top axes of
the plots in Fig.\ \ref{Fig-fc-T}(a,b) computed for spherical particles with
$b=5$ nm and typical YBCO parameters. This scale, however, is very sensitive to
the value of $b$.  For example, the liquid nitrogen temperature $T=77$K
corresponds to the reduced temperature $\tilde{T}\approx 0.041$ in the plot.
For particles with $b=10$nm the same real temperature would correspond to the
much smaller value $\tilde{T}\approx 0.011$.

As we found approximately $f_c\propto \sqrt{n_p}$ at $T=0$, according to the
estimate of Eq.\ (\ref{CrCurrMet_b}) we also expect the relation $f_c\propto
F_p^{3/2}$. This means that, according to Eq.\ (\ref{RenormFp}), we expect the
dependence $[f_c(\tilde{T})/f_c(0)]^{2/3}=1-(\beta_{dp} \tilde{T})^{2/3}$ with
$\beta_{dp}=(1/a_F)\ln(n_{p0}/n_p)$. This is directly verified in Fig.\
\ref{Fig-fc-T}(c)  where we observe the approximate linear dependencies of
$[f_c(\tilde{T})/f_c(0)]^{2/3}$ versus $\tilde{T}^{2/3}$. Moreover, as shown in
the inset, the coefficient $\beta_{dp}$ found from the linear fits for
different pin densities indeed has a logarithmic dependence on $n_p$,
$\beta_{dp}\approx 1.6\ln(0.115/n_p)$. These observations provide justification
for our assumption that the reduction of the typical pin-breaking force is the
main source of thermal suppression of the effective critical force.

We consider now the influence of thermal noise on the long-range behavior of
the line displacements. Figures \ref{Fig-Sl-ul2-T}(a,c) present the evolution
of the structure factor with increasing temperature for the pin density
$n_p=1.33\times 10^{-4}$. We can see that the slope of the small-$q$
dependences does not change indicating that the roughening indices $\zeta_l$
and $\zeta_t$ are temperature independent. However, the coefficient is
significantly reduced for the longitudinal displacements and is enlarged for
the transverse displacements. Correspondingly, as one can see from the line
wandering plots shown in Fig.\ \ref{Fig-Sl-ul2-T}(b,d), the components of
displacements in the critical region have opposite tendencies: the longitudinal
displacements decrease and the transverse displacements increase with
increasing temperature.  As the longitudinal displacements dominate, the lines
become more straight in the critical region. Another important observation is
that the random-force regime $S_{l,t}\propto q^{-4}$ at large $q$ is rapidly
washed out by thermal noise for both components and is replaced by the
isotropic fluctuational line wandering $S_{l,t}=T/q^2$.

\section{Discussion and comparison with experiment \label{Sec-Discussion}}

Using the square-root fit of the pin-density dependence of critical force in
Fig. \ref{Fig-fc-np}, we can restore the $n_{p}$ dependence of the critical
force in real units
\begin{equation}
f_c=1.9 F_p^{3/2}\sqrt{\frac{n_p b}{\varepsilon_1}}
\end{equation}
This coincides with the estimate (\ref{CrCurrMet_b}) which is obtained assuming
that the transverse trapping distance $u_t$ is on the order of the pin size
$b$. This does not contradict too much our numerical results, because our
average values of $u_t$ only slightly exceed the pin size. Note that in our
simulations the pin-breaking force $F_p$ is fixed by interaction with the pin
while in real superconductors for large-size inclusions it is determined by the
in-plane line energy, see Eq.\ (\ref{PinBreakFrc}).  Substituting this
estimate, our result leads to the following estimate for the critical force
\[
f_c=A_c\varepsilon_0\sqrt{n_p b/\gamma},
\]
where, assuming $\xi_c<s$, we estimated $A_c\lesssim 5.4
[\ln(b_z/s)]^{3/2}[\ln(L_t/s)]^{-1/2}$. %

For preliminary comparison with experiment we use results of the recent paper
\onlinecite{PolatPRB11} in which approximately spherical (Y-Gd)$_2$O$_3$
particles with radii $\sim 4$ nm were embedded into the YBCO films. The typical
concentration of particles was $5 \times 10^{16}$cm$^{-3}$. For estimates, we
assume the temperature-dependent London penetration depth as
$\lambda_{ab}=140$nm$/\sqrt{1-(T/T_c)^2}$ with $T_c=90$K and the anisotropy
$\gamma=5$. Such parameters correspond to the reduced concentration
$\tilde{n}_p\approx 1.3\times 10^{-4}$. From the above zero-temperature result,
we estimate $A_c\approx 3.9$ and the critical current $j_c\approx 1.7\times
10^7$ amp/cm$^{2}$ which is close to the low-field experimental value at $5$K,
$2.08\times 10^7$ amp/cm$^{2}$. The reduced temperature can be evaluated as
$\tilde{T} \approx (T/5.3\times10^3K)/[1-(T/T_c)^2]$. At 55 K this gives
$\tilde{T} \approx 0.017$. According to Fig.\ \ref{Fig-fc-T}(b), at this
temperature thermal suppression of the critical current is expected to be
around 50\%. Taking this factor and the temperature dependence of the
parameters into account, we expect a critical current $j_c\approx 5.2\times
10^6$ amp/cm$^{2}$, which again is quite close to the experimental value of
$6.22\times 10^6$ amp/cm$^{2}$.
We can conclude that the our strong-pinning result for the critical
current is in reasonable agreement with experiment.

Our results show that the wandering of dynamically pinned lines is strongly
anisotropic, displacements in the direction of motion are much larger than
displacements in the transverse direction. In principle, experimentally this
can be directly demonstrated using flux visualization techniques, such as a
scanning SQUID or Hall probes. With these techniques individual vortices can
only be resolved at small fields.  Usually statically pinned vortices are
visualized after cooling in fixed magnetic field. Nevertheless, it should also
be possible to visualize shapes of the dynamically-pinned vortices near the
boundary of the Bean profile, which is formed when the magnetic field is
applied after cooling of a superconducting sample in zero field. We expect the
vortex field profiles to be strongly elongated along the direction of motion.
The elongation length along the direction of motion is given by the
longitudinal displacement $u_l$ at a distance on the order of the London
penetration depth. This length is expected to decrease with increasing
temperature.

In conclusion, we developed a quantitative description of individual vortex
lines pinned by an array of nanoparticles. We found that the critical force
grows roughly as the square root of the pin density. This result is consistent
with qualitative estimates assuming that for our pin-density range trapping
events mostly occur as a result of direct collisions with pinning centers. The
apparent critical force is strongly suppressed by thermal noise. The relative
suppression reduces with increasing pin density. The configurations of pinned
lines are strongly anisotropic, displacements in the drive directions are much
larger than those in the transverse direction. The displacement anisotropy is
rapidly reduced by thermal noise mostly due to the rapid reduction of the
longitudinal displacements. This leads to straightening of the lines in the
critical region. Analyzing the behavior of the structure factors at small wave
vectors, we found that the roughening index for the longitudinal displacements
exceeds one. This means that the local stresses in the critical region increase
with the line length indicating a breakdown of the elastic description in the
thermodynamic limit. The behavior of the structure factor at large wave vectors
for both directions is typical for the displacement induced by random force. At
finite temperatures this random-force regime is rapidly replaced by the thermal
displacements.

\begin{acknowledgments}
The authors would like to thank J. R. Thompson for very helpful
discussions and for providing unpublished parameters of the films
studied in Ref. \onlinecite{PolatPRB11}. The authors also
acknowledge useful discussions with D. K. Christen, L. Civale, and
V. B. Geshkenbein.
A.E.K. was supported by UChicago Argonne, LLC, operator of Argonne National
Laboratory, a U.S. Department of Energy Office of Science laboratory, operated
under contract No. DE-AC02-06CH11357. This work was also supported by the
\textquotedblleft Center for Emergent Superconductivity\textquotedblright, an
Energy Frontier Research Center funded by the U.S. Department of Energy, Office
of Science, Office of Basic Energy Sciences under Award Number DE-AC0298CH1088.
A.B.K acknowledges hospitality at Argonne National Laboratory and the support
from CNEA, CONICET under Grant No. PIP11220090100051, and ANPCYT under Grant
No. PICT2007886.
\end{acknowledgments}


\begin{thebibliography}{99}
\bibitem{footnoteLp} This formula is valid only for the case $b\gg
\xi_{ab}$. When $b$ is comparable with $\xi_{ab}$, interpolation
$\mathcal{L}_{p}\approx (1/2)\ln [1+(b/\xi_{ab})^2]$ should be used.
\bibitem{NoteLogLength}The cutoff lengths $r_{\max}$ and $r_{\min}$ depend on
microscopic details. Typically, $r_{\max}$ is determined by the
length scale of line deformation in the $z$ direction, which in our
case is given by the trapping length $L_t$. The lower bound for
$r_{\min}$ is given by the c-axis coherence length $\xi_c$. However,
in layered superconductors more accurately $r_{\min}$ can be
estimated as the typical distance between the pancake vortices in
the neighboring layers devided by the anisotropy factor $\gamma$.
\bibitem{SinglePartNote}Near depinning threshold, the pin-vortex potential
as a function of the pinned-point displacement $u$ can be expanded
as $V(u) = (F_p-F)(u-u_c) - (b/6)(u-u_c)^3$ where $F\approx f L_t$
and $u=u_c$ is the marginal equilibrium point. Below threshold this
yields the barrier $U(F) = [4\sqrt{2}/(3\sqrt{b})](F_p-F)^{3/2}$.
\end{thebibliography}
\end{document}